# Sub-wavelength coherent imaging of periodic samples using a 13.5 nm tabletop high harmonic light source

Dennis F. Gardner[1*], Michael Tanksalvala[1], Elisabeth R. Shanblatt[1], Xiaoshi Zhang[2], Benjamin R. Galloway[1], Christina L. Porter[1], Robert Karl Jr.[1], Charles Bevis[1], Daniel E. Adams[1], Henry C. Kapteyn[1,2], Margaret M. Murnane[1,2] and Giulia F. Mancini[1]

[1]JILA, University of Colorado, 440 UCB, Boulder, CO 80309-0440, USA. [2]Kapteyn-Murnane Laboratories, 1855 S. 57th Court, Boulder, CO 80301, USA. *email: dennis.gardner@colorado.edu

**Coherent diffractive imaging is unique as the only route for achieving diffraction-limited spatial resolution in the extreme ultraviolet and X-ray regions, limited only by the wavelength of the light. Recently, advances in coherent short wavelength light sources, coupled with progress in algorithm development, have significantly enhanced the power of x-ray imaging. However, to date, high-fidelity diffraction imaging of periodic objects has been a challenge because the scattered light is concentrated in isolated peaks. Here, we use tabletop 13.5nm high harmonic beams to make two significant advances. First we demonstrate high-quality imaging of an extended, nearly-periodic sample for the first time. Second, we achieve sub-wavelength spatial resolution (12.6nm) imaging at short wavelengths, also for the first time. The key to both advances is a novel technique called modulus enforced probe, which enables robust, quantitative, reconstructions of periodic objects. This work is important for imaging next generation nano-engineered devices.**

Short wavelength light in the extreme-ultraviolet (EUV) and soft X-ray regions of the spectrum is attractive for high resolution imaging, due to the inherent elemental and chemical-specific contrast in this spectral region resulting from electronic resonances[1–3]. Recent advances in coherent short wavelength light sources, including tabletop high-harmonic generation[4–6] as well as large-scale synchrotron and free electron laser facilities[7], combined with coherent imaging methods, leads to unique new imaging capabilities[8–15]. Coherent diffractive imaging (CDI) techniques[16,17] are particularly attractive for EUV and X-ray imaging because they address critical limitations in short wavelength imaging: most X-ray optics are costly, imperfect, lossy, and cannot reach diffraction-limited spatial resolution[18]. Fortunately, using CDI it is now possible to achieve diffraction-limited spatial resolution at short wavelengths for the first time. Moreover, CDI is the most photon efficient form of imaging because there are no optics between the sample and the detector[19].

In CDI, the amplitude of the scattered light from an object is directly collected on the detector, without the need for any image-forming optics. Although the phase of the light is lost by recording only the diffracted intensity, this can be recovered computationally using iterative phase retrieval algorithms[20–26]. Once the phase of the light is known, the diffracted light can be computationally propagated to the sample plane to retrieve a complex-valued image of the object. For the algorithm to converge to a unique solution, the diffraction intensity must be adequately sampled at the spatial Nyquist frequency in order to regain phase information in the complex-valued diffraction signal[27]. Initial implementations of CDI used mostly pinhole-like test samples that permitted the use of an isolation constraint on the object itself, facilitating phase retrieval[17,28]. However, because the object was isolated, these approaches to CDI could not easily be used to image extended objects of relevance to material science and other applications. Furthermore, these approaches cannot separate the illumination from the object.

More recently, a particularly powerful approach to CDI known as ptychography[29–33] has made it possible to image extended objects by acquiring multiple diffraction patterns from overlapping areas of the sample. Redundant information collected during the scan is harnessed to reconstruct the amplitude and the phase of the sample, as well as the amplitude and phase of the illuminating wave (i.e. the probe)[31,34]. Using ptychography, the isolation constraint is applied directly to the illumination beam instead of the sample, thus enabling diffraction-limited full-field imaging of complex extended objects[35,36]. For non-periodic samples, ptychography can reliably solve for the sample and illumination simultaneously. However, periodic samples pose a particular challenge due to the lack of diversity in the diffraction patterns, resulting in poor convergence of the phase retrieval algorithm[16,17,27,31,37,38]. Moreover, the spatial resolution achieved to date was limited to 1.3$x$ the wavelength ($\lambda$) for real-world objects under good reconstruction conditions[36].

In this work, we achieve a sub-wavelength resolution of 0.9$\lambda$ of an extended sample in the EUV region for the first time using a tabletop 13.5 nm high harmonic light source. We also demonstrate high fidelity, full-field, quantitative imaging of near-periodic objects for the first time. The key to achieving high fidelity imaging of periodic samples is a novel technique we termed Modulus Enforced Probe (MEP). MEP is based on collecting a single image of the un-scattered direct beam on the detector. This additional measurement is then used within a novel algorithm based on the extended ptychographic iterative engine (ePIE)[34] to highly constrain the guess for the illumination beam. Crucially, because the total power in the input beam is measured by recording the direct beam, this method allows *quantitative*[39] amplitude and phase information retrieval for both the sample and illumination. The MEP technique also minimizes reconstruction artifacts due to cross-talk between the sample and the illumination, and allows for a much faster convergence and improved robustness of the algorithm, even in the presence of noise. This experimental demonstration is, to our knowledge, the best wavelength-to-resolution ratio for any full-field, non-isolated sample, CDI-based microscope. Finally, the 13.5 nm wavelength and the ability to reliably image near-periodic objects are technologically relevant in support of the development of next-generation EUV lithography, nanoelectronics, data storage, and self-assembled nanostructures as well as functional imaging of nano-enhanced devices.

**Experimental set-up and Modulus Enforced Probe methodology**

We generated bright, phase-matched, high harmonic beams (HHG)[4–6,40] by focusing an ultrafast 800 nm laser into a Helium-filled waveguide at 1 atm pressure (KMLabs XUUS 4.0). Most of the residual laser light is removed by a rejecter optic placed at an angle near glancing incidence, followed by one 600 nm thick Zr filter. A single harmonic order at a wavelength of 13.5 nm ± 0.2 nm is then selected and focused onto the sample using a pair of Si/Mo multilayer mirrors (Fig. 1). The sample, a zone plate with 150 nm-thick polymethylmethacrylate (PMMA) rings on a 50 nm silicon nitride window, was aligned perpendicular to the beam and placed at the focus (≈2 μm diameter, see Fig. 5). A scanning electron microscope (SEM) image of the ZP sample is shown in the inset of Fig. 1.

The sample was scanned in the focus of the EUV beam in a rectilinear pattern with 121 (11 x 11) positions with nominally 0.88 µm between adjacent positions. To prevent periodic artifacts being introduced by the scan grid itself, a random offset of ±20% of the step size is added to each scan position[41]. At each scan position, light scattered by, and transmitted through, the sample was detected on a back-illuminated CCD placed 22.6 mm ± 0.2 mm from the sample. As shown in Fig. 1, stray HHG light that bypassed the multilayer mirrors is also detected as background light on the detector. To maximize the NA of the microscope, the sample was scanned in the top-right region, such that the strongest diffracted light scattered along the diagonal of the detector. The diffracted light was detected at an NA of 0.54. At this NA, the Abbe diffraction limited resolution, ∆r, is given by:

$$\Delta r = \frac{\lambda}{2\,NA} = 12.5\ nm\ \pm\ 0.2\ nm. \quad (1)$$

Using our novel Modulus Enforced Probe (MEP) methodology, the sample is then moved out of the EUV beam path, and a single image of the direct, un-scattered, HHG illumination is collected at the detector (Fig. S1, supplementary information). This procedure can be carried out before or after the acquisition of the ptychographic dataset. In an alternative approach, MEP can also be applied by separating the DC (un-diffracted beam) from the rest of the diffraction pattern. This is especially effective when it is not practical to collect the direct beam on the detector - for example due to a limited range of scanning stages, or samples that are too large to be removed from the beam path. We used the latter approach in our experiment.

The undiffracted beam was calculated by applying a 30% threshold to the scatter pattern, isolating the DC peak, which served as the MEP constraint (see SI). In order to quantify the flux in the input beam, we statistically averaged several CCD measurements of the un-diffracted beam. In a reflection geometry, (rather than the transmission geometry presented in this paper) a measurement of the reflected beam off of a smooth surface can be used as the un-diffracted beam. If the reflectance of the reference surface is known, then the un-diffracted beam data can be scaled appropriately to yield the quantitative reflectivity of the sample[39].

**Modulus Enforced Probe algorithm implementation**

Figure 2 displays the general working principle of the MEP algorithm. The MEP ptychographic experiment is composed of two data-taking steps. In a first stage, the ptychographic dataset is collected by step-wise scanning of the sample in a coherent EUV beam. In the second step, the direct beam is collected at the detector by moving the sample out of the way of the incident beam. The novel probe constraint, which utilizes the measured probe intensity on the detector, $\mathbb{P}(\boldsymbol{u})$, where $\boldsymbol{u}$ is the spatial frequency coordinate vector on the detector plane, can be implemented in many phase-diverse CDI techniques[31,32,34,42].

A flowchart of how the MEP constraint is employed within the ptychography algorithm at each iteration, *j*, after updating the object and probe, is given in Fig. 2. The updated probe amplitude is propagated to the detector plane, forming a guess of the probe on the detector given by $\mathcal{P}_{Gj}(\boldsymbol{u}) = \mathcal{F}[P_{j+1}(\boldsymbol{r})]$, where $P_{j+1}(\boldsymbol{r})$ is the updated probe, $\boldsymbol{r}$ is the spatial coordinate vector at the sample plane, $\mathcal{F}$ is a propagator from the sample plane to the detector plane, and $\mathcal{P}_{Gj}(\boldsymbol{u})$ is a guess of the probe at the detector plane. Here, we apply a modulus constraint to the probe guess enforcing the probe measurement. This gives:

$$\mathcal{P}_{Mj}(\boldsymbol{u}) = \sqrt{\mathbb{P}(\boldsymbol{u})}\frac{\mathcal{P}_{Gj}(\boldsymbol{u})}{|\mathcal{P}_{Gj}(\boldsymbol{u})|^2}. \quad (3)$$

This modulus-constrained probe guess is propagated back to the sample plane, $\mathcal{F}^{-1}$, forming a further updated probe guess:

$$P'_{Gj+1}(\boldsymbol{r}) = \mathcal{F}^{-1}[\mathcal{P}_{Mj}(\boldsymbol{u})] \qquad (4)$$

which is consistent with the measured probe intensity. The new probe guess, $P'_{Gj+1}$ is now fed back into the algorithm.

This additional constraint to the probe is analogous to the error-reduction algorithm as described in by Fienup[20], but instead of using error-reduction to solve for the object, it is used within the ptychography algorithm to further constrain the probe guess. The MEP constraint provides additional information for the ptychography algorithm, thereby improving the convergence speed (see SI).

**Sub-wavelength coherent diffractive imaging of extended near-periodic objects**

Reconstructed intensity images of the zone plate sample, obtained with the ePIE ptychography algorithm[34] *with and without* the novel MEP constraint, are displayed in Fig. 3(b, c) respectively. The high-fidelity ptychographic reconstruction obtained with MEP (Fig. 3b) is compared to the SEM shown in Fig. 3a. The SEM image shows an inverted contrast with respect to Fig. 3b, due to the reflection mode of the SEM, as opposed to our ptychographic CDI transmission microscope. The full, reconstructed images, as well as the retrieved phase images are given in the supplementary information (SI).

Although the width of the rings in the zone plate have a radial dependence proportional to *1/r*, with *r* = 40 μm for the radius of the outer-most ring, the PMMA rings within the 5.65 μm *x* 5.65 μm area shown in Fig. 3(b, c) constitute a periodic arrangement of lines. The periodicity of these features represents a major challenge in traditional ptychographic CDI due to the lack of diversity in the diffraction patterns, leading to poor convergence of the phase retrieval algorithms. As a result, when the MEP methodology is *not* applied, ringing artifacts indicated by stripes in the image are seen in the reconstruction (Fig. 3c).

The highlighted regions of Fig. 3b and Fig. 3c show the periodic arrangement of 8 PMMA lines with 110 nm ± 10 nm width spaced by 90 nm ± 10 nm of $Si_3N_4$. The high-fidelity image obtained with MEP (blue box inset to Fig. 3b) is compared to the same region imaged without applying MEP (orange box inset to Fig. 3c), and to the image obtained with the SEM (black box inset to Fig. 3a). A profile across a zone plate PMMA feature was selected (blue line, inset Fig. 3b) for the reconstruction obtained with the MEP constraint. The corresponding lineout is shown in Fig. 4a by blue circle data points and was fitted to an error function. The 10% - 90% width of the fit (r-square better than 0.97) is better than our reconstruction pixel size, dx, given by

$$dx = \frac{\lambda z}{N\, px} = 12.6\; nm \pm 0.2\; nm, \quad (2)$$

where λ is the wavelength, *z* is the distance from the CCD to the sample, *N* is the number of pixels on the detector, and *px* is the effective pixel size with on-chip binning. The agreement between the 12.6 nm resolution supported by the lineout of the zone plate feature and the 12.5 nm Abbe limit demonstrates that the algorithm was able to converge to a solution limited only by the NA of the data, and demonstrating sub-wavelength EUV imaging for the first time.

For the analysis of the reconstruction *without* the MEP constraint, 10 parallel and adjacent profiles were averaged (orange line, inset of Fig. 3c), due to the presence of artifacts in the image. The corresponding lineout is shown by the orange square data points in Fig. 4a and fitted to an error function (r-squared better than 0.98). The 10% - 90% values of the fit support a 60 nm resolution without the MEP constraint.

In addition to the high-spatial resolution of 0.9λ, we can also obtain topographical information from the phase images. The phase information from the inset of Fig. 3b, is used to generate a three-dimensional rendering of the zone plate in Fig. 4b. The reported values of PMMA[43] were used to calculate the height of the zone plate features to be 180 nm ± 40 nm. The phase images of the zone plate are shown in the SI (Fig. S4 and S5b).

An important implication of the enforcement of the total power of the illumination during the image reconstruction is that the reconstructed amplitude image encodes the absolute transmission from the sample, as shown in Fig. 3b. This capability enables quantitative imaging, the importance of which has been recently demonstrated with EUV RAPTR-CDI on buried layered structures[39].

Most importantly, the MEP constraint brings unique capabilities of enabling reconstructions of periodic objects by minimizing the cross-talk between object and probe, as demonstrated by the smooth and continuous amplitude and phase retrieved for the illumination (Fig. 5 a, b). Without the MEP constraint (Fig. 5 c, d), the illumination exhibits ringing artifacts. Cross-talk between the illumination and object (Fig. 3c) is present because the algorithm transfers power between the object and probe, which results in unreliable transmission values for the sample. Furthermore, the ptychography algorithm converges faster with the MEP constraint, especially in the presence of a poor initial probe guess, and the MEP constraint improves robustness in the presence of background. A detailed discussion on the MEP constraint effects on the algorithm convergence and the effects of Gaussian and Poisson noise are reported in the SI.

We note that previous work suggested sub-wavelength resolution in the EUV from a toy, pinhole-like sample, with algorithms that use a low degree of data redundancy and are not capable of imaging extended or periodic objects[28]. Even for relatively thin samples, these methods do not produce reliable resolution limits especially due to the fact that modulations in the exit surface wave (ESW) can lead to artificially high spatial frequency content not found in the object. Furthermore the estimated spatial resolution was based solely on the phase-retrieval transfer function (PRTF) and did not include more direct measurements like the knife-edge test.

**Conclusions**

We demonstrated 0.9λ spatial resolution imaging of a periodic extended object with 13.5 nm EUV light from a high-harmonic source, by applying a novel modulus enforced probe technique. Without the MEP constraint, the reconstructed image is characterized by lower fidelity (60 nm resolution) and by cross-talk between the object and the probe. With a modulus enforced probe, we achieve 12.6 nm ± 0.2 nm spatial resolution imaging, which agrees with the theoretical resolution of 12.5 nm ± 0.2 nm corresponding to the numerical aperture. This represents, to our knowledge, a record wavelength-to-resolution ratio for full-field imaging of an extended object with any EUV/x-ray source. Moreover, our work presents the first high-fidelity image of a periodic sample using a lensless imaging technique.

Ongoing research that requires more advanced imaging techniques such as multiplexed wavelength imaging[11,44,45], undersampling[46,47], and thick samples[48,49] will all benefit from the MEP constraint. In all these cases, the ptychography algorithm is tasked with solving for more than just a single sample and illumination. The additional information introduced by the MEP constraint will reduce the number of unknowns in the algorithm, likely helping with convergence and stability.

Future studies can employ the EUV and x-ray spectral region for chemical and elemental specificity. Furthermore, illumination with shorter wavelengths is possible using either tabletop high-harmonic sources or large-scale synchrotrons and X-ray free electron lasers, which in the future can enable nanometer or even atomic scale resolution of a broad range of next generation nanoelectronics, data storage and nano-engineered systems.

**Methods**

**Experimental layout**: The driving laser is a KMLabs Dragon centered at a wavelength of 785 nm, with 2 mJ pulse energy, 23 fs pulse duration, at a repetition rate of 3 kHz. The laser is focused and coupled into a 150 μm diameter capillary, filled with 500 Torr of Helium, which is carefully engineered to phase-match the generation of harmonics at 13.5 nm. The second multilayer mirror has a radius-of-curvature of 100 mm. The angle-of-incidence on the mirrors is estimated to be 2° ± 0.5° from normal. The beam size at the focus was measured using a knife-edge method to be approximately 2 μm ± 0.5 μm in diameter. The EUV light was detected on an Andor iKon with an array of 2048 x 2048 square pixels (side length of 13.5 μm). The collected diffraction patterns were binned by 2 on-chip and cropped to 900 x 900 pixels. At each scan position, two accumulations were taken with 4.25s exposures at 1MHz pixel readout rate.

**Image reconstructions**: The reconstructions were done in four stages with the ePIE algorithm[34] as described in Ref. [34]. In one set of reconstructions, we modified the ePIE algorithm and implemented our MEP constraint. In the first two stages a binary mask was used to block the stray background light. In the first stage, we used a calculated probe through the beamline and an object guess of unity. The algorithm performed 950 iterations with probe updates allowed after 100 iterations. In the second stage, we re-initialized the object guess to unity and fed in the probe from the end of stage 1. Stage 2 was allowed to run for 700 iterations with updates allowed to the probe after iteration 100. In the third stage, we used the probe from the end of stage 2 and the object guess was re-initialized to unity. At this stage, we let the algorithm fill in data in the masked-off background region. In other words, the modulus constraint was not enforced in the stray background region. Stage 3 ran for 10,000 iterations with probe updates allowed between iterations 2,000 through 9,000. Finally, in stage 4, we fed in both the object and probe from the end of stage 3. The algorithm was allowed to continue to extrapolate values and position correction was performed as described in Ref. [50] between iterations 2 – 9000[50]. The probe was allowed to update between iterations 2,000 - 9,000. The final reconstructions, and retrieved probes, obtained after 10,000 iterations, are shown in Fig. 3, Fig. 5 and in the SI. For numerical aperture values higher than the 0.54 presented in this work, the curvature of the diffraction on the Ewald sphere must be taken into account before carrying out the image reconstruction.

**2+1D phase reconstruction**: The intensity images of the zone plate were used to create two binary masks: a mask for the PMMA features and a mask for the $Si_3N_4$ substrate. A threshold of 1/3 of the theoretical transmission of $Si_3N_4$[43] was used to separate the materials. The phase of the substrate was fitted to a 2$^{nd}$ order plane. The fitted phase plane was subtracted from the entire phase image. Values less than –π were wrapped around to lie between –π and π. Within the PMMA mask, values less than 3 times the standard deviation of the substrate phase were increased by π to unwrap the phase. The phase values were converted to height using the reported index of refraction values in Ref. [43] and a wavelength of 13.5 nm. The rendered image is interpolated onto a 4x larger grid and the vertical scale is scaled by 1/5.

**References**

1. Shapiro, D. A. *et al.* Chemical composition mapping with nanometre resolution by soft X-ray microscopy. *Nat. Photonics* **8,** 765–769 (2014).
2. Donnelly, C. *et al.* Element-Specific X-Ray Phase Tomography of 3D Structures at the Nanoscale. *Phys. Rev. Lett.* **114,** 1–5 (2015).
3. Beckers, M. *et al.* Chemical Contrast in Soft X-Ray Ptychography. *Phys. Rev. Lett.* **107,** 208101 (2011).
4. Rundquist, A. *et al.* Phase-Matched Generation of Coherent Soft X-rays. *Science* **280,** 1412–1415 (1998).
5. Bartels, R. A. *et al.* Generation of spatially coherent light at extreme ultraviolet wavelengths. *Science* **297,** 376–378 (2002).
6. Zhang, X. *et al.* Highly coherent light at 13 nm generated by use of quasi-phase-matched high-harmonic generation. *Opt. Lett.* **29,** 1357 (2004).
7. Ishikawa, T. *et al.* A compact X-ray free-electron laser emitting in the sub-ångström region. *Nat. Photonics* **6,** 540–544 (2012).
8. Miao, J., Ishikawa, T., Robinson, I. K. & Murnane, M. M. Beyond crystallography: Diffractive imaging using coherent x-ray light sources. *Science* **348,** 530–535 (2015).
9. Clark, J. N. *et al.* Ultrafast Three-Dimensional Imaging of Lattice Dynamics in Individual Gold Nanocrystals. *Science* **341,** 56–59 (2013).
10. Leshem, B. *et al.* Direct single-shot phase retrieval from the diffraction pattern of separated objects. *Nat. Commun.* **7,** 10820 (2016).
11. Odstrcil, M. *et al.* Ptychographic imaging with a compact gas–discharge plasma extreme ultraviolet light source. *Opt. Lett.* **40,** 5574–5577 (2015).
12. Pfeifer, M. a, Williams, G. J., Vartanyants, I. a, Harder, R. & Robinson, I. K. Three-dimensional mapping of a deformation field inside a nanocrystal. *Nature* **442,** 63–6 (2006).
13. Dierolf, M. *et al.* Ptychographic X-ray computed tomography at the nanoscale. *Nature* **467,** 436–439 (2010).
14. Roy, S. *et al.* Lensless X-ray imaging in reflection geometry. *Nat. Photonics* **5,** 243–245 (2011).
15. Sun, T., Jiang, Z., Strzalka, J., Ocola, L. & Wang, J. Three-dimensional coherent X-ray surface scattering imaging near total external reflection. *Nat. Photonics* **6,** 586–590 (2012).
16. Sayre, D. Prospects for long wavelength X-Ray microscopy and diffraction. in *Imaging processes and coherence in physics* **112,** 229–235 (1980).
17. Miao, J., Charalambous, P., Kirz, J. & Sayre, D. Extending the methodology of X-ray crystallography to allow imaging of micrometre-sized non-crystalline specimens. *Nature* **400,** 342–344 (1999).
18. Chao, W., Harteneck, B., Liddle, J., Anderson, E. & Attwood, D. Soft X-ray microscopy at a spatial resolution better than 15 nm. *Nature* **435,** 1210–1213 (2005).
19. Huang, X. *et al.* Signal-to-noise and radiation exposure considerations in conventional and diffraction x-ray microscopy. *Opt. Express* **17,** 13541–13553 (2009).
20. Fienup, J. Reconstruction of an object from the modulus of its Fourier transform. *Opt. Lett.* **3,** 27–29 (1978).
21. Fienup, J. R. Phase retrieval algorithms: a comparison. *Appl. Opt.* **21,** 2758–69 (1982).
22. Elser, V. Random projections and the optimization of an algorithm for phase retrieval. *J. Phys. A. Math. Gen.* **36,** 2995–3007 (2003).
23. Marchesini, S. *et al.* X-ray image reconstruction from a diffraction pattern alone. *Phys. Rev. B* **68,** 140101 (2003).

24. Luke, D. R. Relaxed Averaged Alternating Reflections for Diffraction Imaging. *Inverse Probl.* **37,** 13 (2004).
25. Thibault, P. & Guizar-Sicairos, M. Maximum-likelihood refinement for coherent diffractive imaging. *New J. Phys.* **14,** 1–20 (2012).
26. Shechtman, Y. *et al.* Phase Retrieval with Application to Optical Imaging: A contemporary overview. *IEEE Signal Process. Mag.* **32,** 87–109 (2015).
27. Miao, J., Sayre, D. & Chapman, H. N. Phase retrieval from the magnitude of the Fourier transforms of nonperiodic objects. *J. Opt. Soc. Am. A* **15,** 1662 (1998).
28. Zürch, M. *et al.* Real-time and Sub-wavelength Ultrafast Coherent Diffraction Imaging in the Extreme Ultraviolet. *Sci. Rep.* **4,** 1–5 (2014).
29. Rodenburg, J. M. *et al.* Hard-X-ray lensless imaging of extended objects. *Phys. Rev. Lett.* **98,** 1–4 (2007).
30. Rodenburg, J. M. Ptychography and related diffractive imaging methods. *Adv. Imaging Electron Phys.* **150,** 87–184 (2008).
31. Thibault, P. *et al.* High-resolution scanning x-ray diffraction microscopy. *Science* **321,** 379–82 (2008).
32. Guizar-Sicairos, M. & Fienup, J. Phase retrieval with transverse translation diversity: a nonlinear optimization approach. *Opt. Express* **16,** 7264–7278 (2008).
33. Kane, D. J. Method and apparatus for determining wave characteristics from wave phenomena. *US Pat. 6,219,142 B1* (2001).
34. Maiden, A. & Rodenburg, J. An improved ptychographical phase retrieval algorithm for diffractive imaging. *Ultramicroscopy* **109,** 1256–62 (2009).
35. Seaberg, M. *et al.* Tabletop Nanometer Extreme Ultraviolet Imaging in an Extended Reflection Mode using Coherent Fresnel Ptychography. *Optica* **1,** 39–44 (2014).
36. Zhang, B. *et al.* High contrast 3D imaging of surfaces near the wavelength limit using tabletop EUV ptychography. *Ultramicroscopy* **158,** 98–104 (2015).
37. Sayre, D. Some implications of a theorem due to Shannon. *Acta Crystallogr.* **5,** 843–843 (1952).
38. Harada, T., Nakasuji, M., Nagata, Y., Watanabe, T. & Kinoshita, H. Phase Imaging of Extreme-Ultraviolet Mask Using Coherent Extreme-Ultraviolet Scatterometry Microscope. *Jpn. J. Appl. Phys.* **52,** 06GB02 (2013).
39. Shanblatt, E. R. *et al.* Quantitative Chemically-Specific Coherent Diffractive Imaging of Buried Interfaces using a Tabletop EUV Nanoscope. *Nano Lett.* **19,** 5444–5450 (2016).
40. Kapteyn, H. C., Murnane, M. M. & Christov, I. P. Extreme nonlinear optics: Coherent X rays from lasers. *Phys. Today* **58,** 39–44 (2005).
41. Thibault, P., Dierolf, M., Bunk, O., Menzel, A. & Pfeiffer, F. Probe retrieval in ptychographic coherent diffractive imaging. *Ultramicroscopy* **109,** 338–43 (2009).
42. Putkunz, C. T. *et al.* Phase-diverse coherent diffractive imaging: High sensitivity with low dose. *Phys. Rev. Lett.* **106,** 1–4 (2011).
43. Henke, B. L., Gullikson, E. M. & Davis, J. C. X-Ray Interactions: Photoabsorption, Scattering, Transmission, and Reflection at E = 50-30,000 eV, Z = 1-92. *At. Data Nucl. Data Tables* **54,** 181–342 (1993).
44. Thibault, P. & Menzel, A. Reconstructing state mixtures from diffraction measurements. *Nature* **494,** 68–71 (2013).
45. Batey, D. J., Claus, D. & Rodenburg, J. M. Information multiplexing in ptychography. *Ultramicroscopy* **138,** 13–21 (2013).
46. Edo, T. B. *et al.* Sampling in x-ray ptychography. *Phys. Rev. A - At. Mol. Opt. Phys.* **87,** 1–8 (2013).

47. Batey, D. J. *et al.* Reciprocal-space up-sampling from real-space oversampling in x-ray ptychography. *Phys. Rev. A - At. Mol. Opt. Phys.* **89,** 1–5 (2014).
48. Maiden, A. M., Humphry, M. J. & Rodenburg, J. M. Ptychographic transmission microscopy in three dimensions using a multi-slice approach. *J. Opt. Soc. Am. A* **29,** 1606–1614 (2012).
49. Godden, T. M., Suman, R., Humphry, M. J., Rodenburg, J. M. & Maiden, A. M. Ptychographic microscope for three-dimensional imaging. *Opt. Express* **22,** 12513–12523 (2014).
50. Zhang, F. *et al.* Translation position determination in ptychographic coherent diffraction imaging. *Opt. Express* **21,** 13592–13606 (2013).

**Acknowledgements**

DARPA PULSE: W31P4Q-13-1-0015, Gordon and Betty Moore Foundation's EPiQS Initiative through Grant GBMF: NSF MRSEC: DMR-1420620; NSF STC: DMR- 1548924; DOE STTR Grant No. DE-SC0006514; and the following Fellowships: 4538, NSF COSI IGERT: 0801680, NSF GRFP: 1144083, Katharine Burr Blodgett Fellowship, Ford Fellowship, Swiss National Science Foundation grant No.P2ELP2_158887, NDSEG fellowship.

**Author contributions**

HCK and MMM conceived of the experiment. All authors designed aspects of the experiment, performed the research and wrote the paper. DFG and GFM characterized the source and collected the data sets. DFG did the reconstructions and data analysis. GFM did the SEM imaging of the zone plate. DEA, DFG, and MT did the probe enforcement simulations. XZ, HCK, MMM and BRG designed the HHG source.

**Additional Information**

Supplementary information: (1) Modulus Enforced Probe Simulations. (2) Phase images with and without the novel MEP constraint. (3) Full ptychographic reconstructed image. (4) Transmission values obtained for the Fresnel Zone Plate (5) Assignment of CCD counts to the MEP un-diffracted beam.

E.S., C.P., M.T., D.G., D.A., G.M., M.M. and H.K. have submitted a patent disclosure based on this work

M.M. and H.K. are partial owners of Kapteyn-Murnane Laboratories Inc. who manufactured the ultrafast laser and EUV source.

# Figures

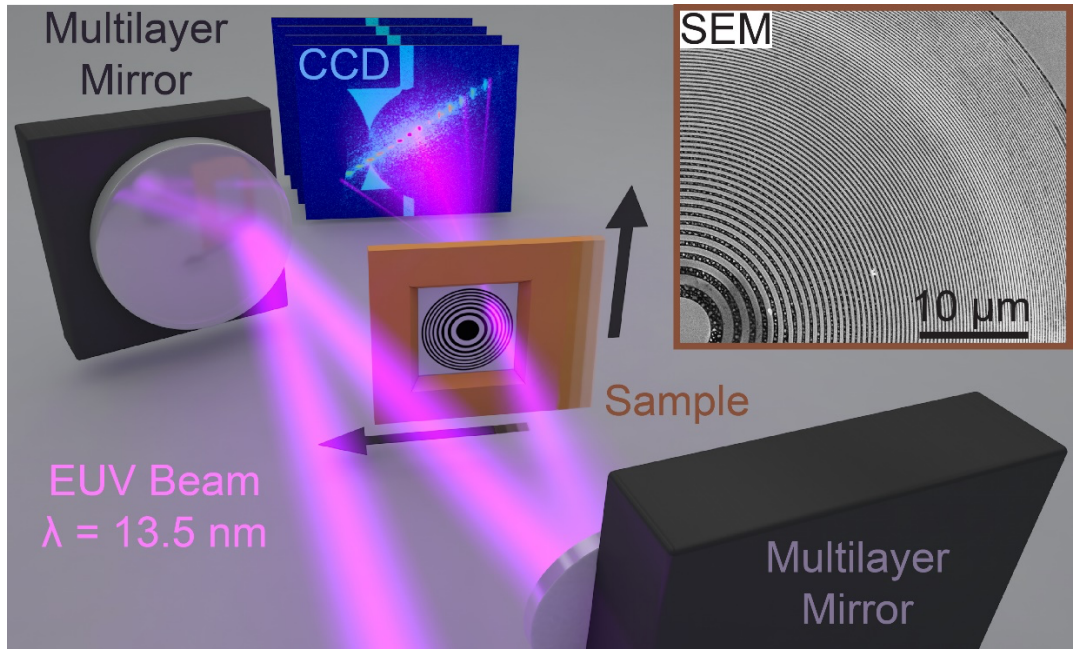

**Figure 1 | Modulus Enforced Probe X-ray microscopy with 13.5 nm extreme ultraviolet light from a tabletop high-harmonic source.** Multilayer mirrors are used to select and focus a single harmonic onto the zone plate sample. The light scattered from the sample, as well as some background light, is collected on an EUV sensitive charged-coupled device (CCD). The inset shows a Scanning Electron Microscope (SEM) image of the sample. The sample is raster scanned and the diffraction from each position is collected.

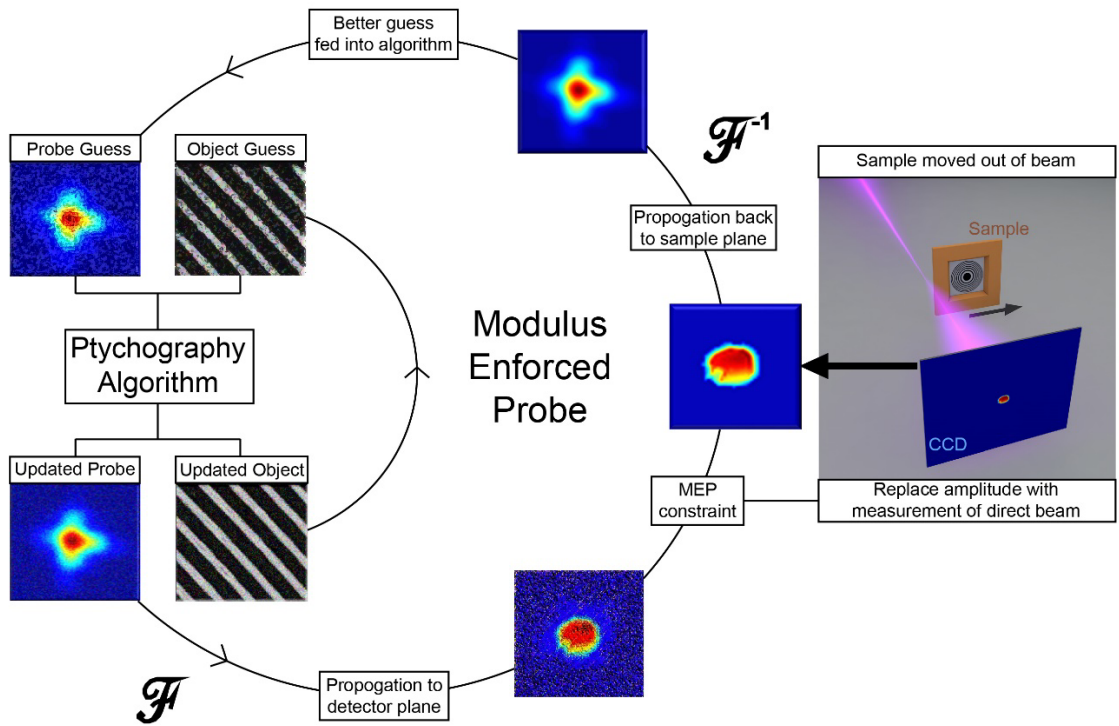

**Figure 2 | Schematic layout of the Modulus Enforced Probe constraint within the ptychography algorithm.** The ptychography algorithm starts with a guess of the object and probe, and then uses the set of diffraction measurements and the overlap of adjacent positions to iteratively update the object and probe guesses. The updated probe is further constrained with the measurement of the probe on the detector. This MEP constrained probe is fed back into the ptychography algorithm yielding faster convergence towards a unique solution.

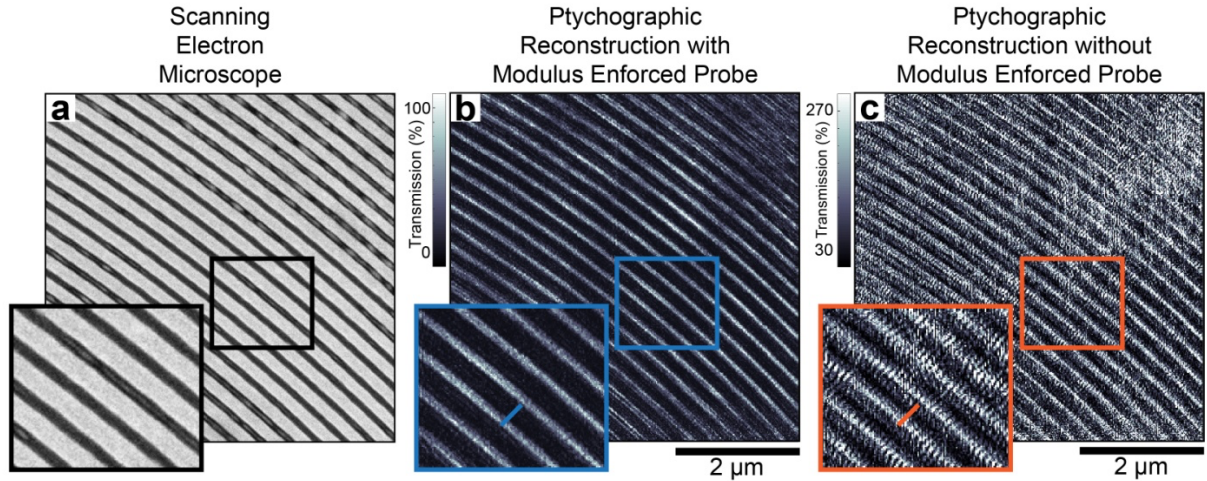

**Figure 3 | Record 0.9λ sub-wavelength resolution full-field imaging using 13.5 nm light and Modulus Enforced Probe. a**, Scanning Electron Microscope (SEM) image of the zone plate sample. **b**, Ptychographic reconstruction of the sample, in the same region, obtained with the novel Modulus Enforced Probe (MEP) constraint. The MEP constraint allows the algorithm to solve for the absolute transmissivity of the sample, which enables unique quantitative imaging capabilities. The contrast of the transmission EUV image is inverted with respect to SEM image in (**a**), obtained in reflection mode. **c**, Ptychographic reconstruction *without* the MEP constraint. Without the additional constraint, unphysical transmission values are obtained.

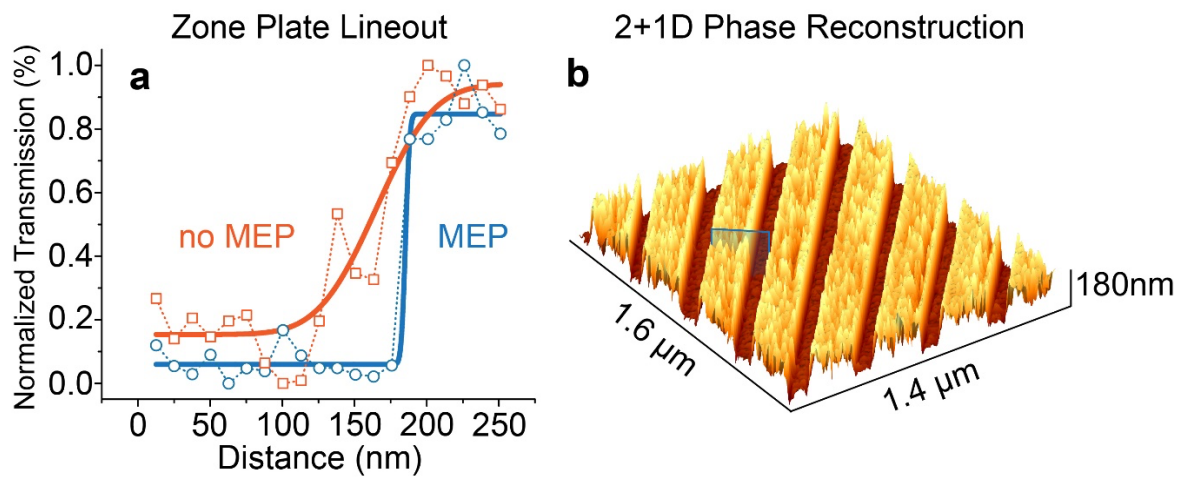

**Figure 4 | Zone plate lineout and height map. a**, Profile of a ZP feature with (blue) and without (orange) the MEP constraint. The data points of the profile are fitted to an error function. The 10% - 90% width from the fit of the data with the MEP constraint supports 12.6 nm sub-wavelength resolution. **b**, Height map of the ZP sample, obtained from the phase using the MEP constraint. The vertical scale bar has been scaled by 1/5. The blue plane shows the location of the lineout in (**a**).

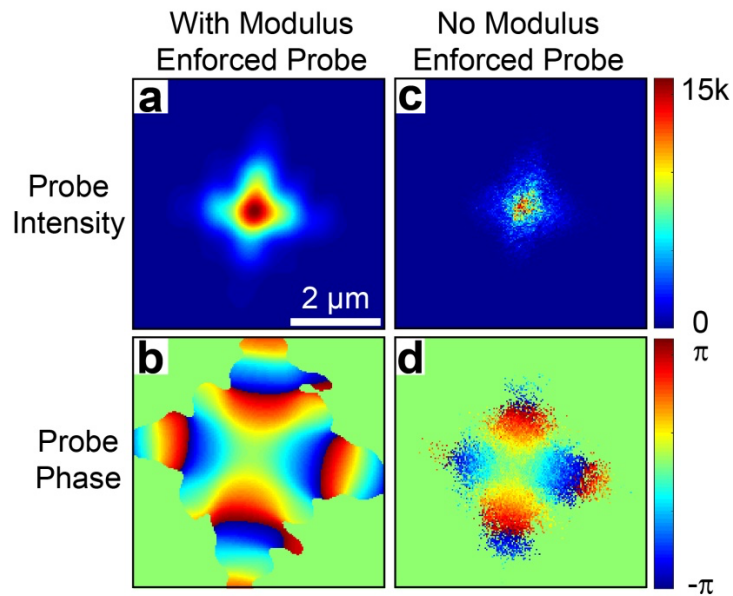

**Figure 5 | Minimization of artifacts in the reconstructed illumination with probe enforcement. a**, Intensity and **b,** phase of the EUV illumination at the sample plane with the Modulus Enforced Probe (MEP) constraint. The intensity of the illumination is shown in detector counts. **c**, Intensity and **d**, phase of the illumination *without* the MEP constraint. Without the MEP constraint there is cross-talk between the illumination and sample leading to poor decoupling. The scale bar in (**a**) is common to all the panels.